\documentclass[11pt]{article}
\usepackage{hyperref}
\usepackage{cite}
\usepackage{xcolor}
\usepackage{amsmath}
\usepackage{amssymb}
\usepackage{footnote}
\makesavenoteenv{minipage}
\newcommand{\nl}{\\ \nonumber && }
\renewcommand{\title}[1]{%
    \bigskip%
    \begin{center}%
    \Large\bf #1%
    \end{center}%
    \vskip .2in}

\renewcommand{\author}[1]{%
    {\begin{center}
    #1
    \end{center}}}
\newcommand{\address}[1]{\vspace{-1.7em}\vspace{0pt}
    {\begin{center}
    \it #1
    \end{center}}}

\begin{document}

\begin{titlepage}
\title{BRST symmetry for Regge-Teitelboim based minisuperspace models}
\author
{ 
Sudhaker Upadhyay $\,^{\rm a,b}$, and
Biswajit Paul    $\,^{\rm c, d}$}
 
\address{$^{\rm a}$Centre for Theoretical Studies, 
Indian Institute of Technology Kharagpur,  Kharagpur-721302, WB, India}

\address{$^{\rm c}$Department of Physics, National Institute of Technology  Agartala, Tripura, India - 799046  }
 
\address{$^{\rm b}$\tt sudhakerupadhyay@gmail.com}

\address{$^{\rm d}$\tt  biswajit.thep@gmail.com}
  
\begin{abstract}
The Einstein-Hilbert action in the context of Higher derivative theories  is considered for finding out their BRST symmetries. Being a constraint system, the model is transformed in the  minisuperspace language with the FRLW background and the gauge symmetries are explored. Exploiting the first order formalism developed by Banerjee et. al. the   diffeomorphism symmetry is extracted. From the general form of the gauge transformations of the field, the analogous BRST transformations are calculated. The effective Lagrangian is constructed by considering two gauge fixing conditions. Further, the
BRST (conserved) charge is computed which plays an important role in defining 
the physical states from the total Hilbert space of states. The finite field dependent BRST (FFBRST) formulation is also studied in this context where the Jacobian for functional measure is illustrated specifically. 
\end{abstract}
\end{titlepage} 
%\newpage

\section{Introduction}

 In field theory, higher derivative terms  are often considered to describe renormalisation of the corresponding theory. In \cite{stelle} Stelle showed that higher derivative gravity models can be renormalised.  Being a simple model, the Einstein -Hilbert(EH) action, when expressed with respect to the metric components shows the higher derivative nature. A careful examination of the  action shows that one can take out a total derivative term which is known as the Gibbons-Hawking term \cite{gibbon}.  This makes the Einstein-Hilbert theories a first order one. Although, in usual theories the surface terms are neglected, but in gravitation they serve as important information about entropy of the system. Therefore, removing the surface term trivially for the gravity models may lead us to losing important information about the thermodynamics of the system.  This motivates us to keep the total derivative term intact to the action thereby making it a higher derivative theory. 
 
 In the minisuperspace version one considers finite degrees of freedom of a model to study the full theory.  With  Friedmann-Rebertson-Lamaitre-Walker (FRLW) background we construct the minisuperspace version of   the EH action which is analogous to  the minisuperspace version of the  Regge-Teitelboim (RT) model \cite{regge}.  Hamiltonian analysis of this minisuperspace model has  already been performed in different ways and  quantization was further explored in \cite{davidson1,  karasik,BMP_RT,cordero1, cordero2}. Particularly, in \cite{BMP_RT} the same model with the higher derivative term was analysed without neglecting the surface term. There the authors have explored the reparammetrisation symmetries of the  model exploiting the first order formalism developed earlier in \cite{BMP}. To our knowledge, the BRST analysis for this model is still lacking in the literature which can be important from the view of quantization of the gravity models. Therefore, this paper will serve the purpose visualizing the role of BRST symmetries in the   cosmological models which will serve important for the purely higher derivative cosmological models. 
 
 The BRST symmetry and the associated concept of BRST cohomology provide the most used covariant quantization method for constrained  systems such as gauge \cite{BPU}, string  and gravity theories  \cite{ht,wei}.  There
appear geometrical constraints, for instance in topological solitons such as 
nonlinear sigma model, CP(N) model, Skyrmion model and chiral bag model,  which can be rigorously treated in the Hamiltonian
quantization scheme  in extended phase spaces \cite{hon}. The extended phase space includes ghost and antighost fields of the theories. This symmetry guarantees the quantization, renormalizability, unitarity and 
other aspects of the first-class constrained  
theories. 
The derivation
of Slavnov-Taylor identities  utilize the  BRST symmetry 
transformation.
 The generalization of BRST transformation   by 
  allowing the parameter  finite and field-dependent, known as  FFBRST  symmetry \cite{jm},  are studied extensively for diverse area \cite{ffb,smm,sudd,rsu,fsm,sb}. 

In the present paper, the minisuperspace version of the EH action is considered to study its BRST symmetries.   For that we first perform the Hamiltonian  formulation exploiting the first order formalism developed in \cite{BMP}.  Dirac's method  for  constraint analysis is followed  to find out the full  constraint structure of the model \cite{dirac, hanson, ht}. The theory is found to have only one primary first class constraint  and one secondary first class constraint. Existence of the  primary first class constraint is in conformity with  one independent gauge symmetry which is in the course of calculation identified as reparametrisation invariance. To canonically quantize the model two gauges were proposed                                                                                                                                                                                                                                                                                                                                                                                                                                                                                                                                                                                                                                                                                                                                                                                                                                                                                                                                                                                                                                                                                                                                                                                                                                                                                                                                                             in \cite{BMP_RT}.  We considered the same gauge fixing conditions to find out their implications in the BRST symmetries.   We found out the nilpotent symmetries which are in line with the standard characteristics of  the BRST symmetries. Also the BRST charge is calculated using Noether charge definition. Further the BRST symmetry is generalized by integrating the transformation parameter after making it field-dependent. By doing so 
we obtain a new symmetry of the theory.
The main feature of this symmetry is that this leads to a non-trivial
Jacobian for functional measure under change of variables, however the effective action remains
unchanged.
We compute the Jacobian systematically for an arbitrary and for some specific choices.
  We observe that such analysis amounts certain change in the effective action.
  Since this change appears in the BRST exact part, so it is not harmful at all for the theory on physical ground.

The paper is organized as follows. In section II we give a very brief introduction to the RT model. In the section III we study the gauge symmetries using the first order formalism. Section IV is our main contribution which we devoted to explore the BRST symmetries.  In this section we calculated the BRST transformation using the same gauges as considered in \cite{BMP_RT}. Also, we calculated the BRST charge  and the Jacobian of the path integral under FFBRST symmetry in section V.  Finally,  we conclude in section VI.

\section{Minisuperspace version of the Einstein-Hilbert action}
In this section we convert the Einstein-Hilbert action to the minisuperspace version keeping the FRLW background. As mentioned earlier, this has a direct representation as the  Regge--Teitelboim model  [see \cite{cordero1, BMP_RT}]. As the Universe has degrees of freedom, limiting the degrees of freedom can make life easier. So we reduce the dimension of  spacetime which is the underlying concept for minisuperspace. Let us take a d-dimensional brane $\Sigma$  evolving in a $N$ dimensional bulk spacetime with fixed Minkowski metric $\eta_{\mu\nu}$. The world volume swept out by the brane is a $d+1$ dimensional manifold $m$ defined by the embedding $x^\mu = X^\mu(\xi^a)$ where $x^\mu$ are the local coordinates of the background spacetime and $\xi^a$ are local coordinates for $m$. The theory is given by the action functional
\begin{equation}
S[X] = \int_{m} d^{d+1} \xi  \sqrt{-g} (\frac{\beta}{2}\mathcal{R} - \Lambda). \label{main_lag}
\end{equation}
Clearly, in this action integral $\beta$ is should be having the dimension $[L]^{1-d}$. Here $g$ is the determinant of the metric, $\Lambda$ is the cosmological constant and  $\mathcal{R}$, the Ricci scalar. 

 Next, we take a  homogeneous and isotropic universe in 4-dimensions and embed it  locally  in a 5-dimensional Minkowski spacetime.  The reason behind taking this choice of  dimension is worth mentioning.  Algebraically, the embedding of a 4 dimensional space in a 5D Minkowski space is correct but on the physical ground it has a limitation only to the vacuum solutions  with cosmological constant. This is in line with the findings of Kasner who first pointed out that most curved  solutions in 4D can not be embedded in 5 dimensions\cite{kasner}. On the other hand, we can embed a all 4 dimensions with vacuum solutions to 6 or higher dimensions. This helps us to understand the dynamics of  4D space with the help of  extra dimension.   
 %%\begin{equation}
% ds^{2} = - dt^{2} + da^{2} + a^{2}d\Omega_{3}^{2},
%\end{equation}  
%where $d\Omega_{3}^{2}$ is the metric for unit 3 sphere.
 %The system is parametrise by $\tau$. Considering spherical symmetry only $t(\tau)$ and $a(\tau)$ are the dynamical variables.

 Now we decompose the Lagrangian in the action (\ref{main_lag}) in the ADM formalism \cite{deser}. In (4+1) dimensions if  we take the normal vector ($N=\sqrt{\dot{t}^{2}-\dot{a}^{2}}$ is the lapse function and the over dot `$^.$' is derivative with respect to $\tau$)
 \begin{equation}
 n_{\mu} = \frac{1}{N}(-\dot{a}, \dot{t}, 0,  0,0),
 \label{normal_vec}
 \end{equation}
   the induced metric on the world volume is,
  \begin{equation}
  ds^{2} = -N^{2} d\tau^{2} + a^{2} d \Omega_{2}^{2}.
  \label{metric}
\end{equation}
We will consider the above metric for the analysis of the system. The system is parametrise by $\tau$. For a spherically symmetric system only  $t(\tau)$ and $a(\tau)$ are the  dynamical variables and other variables can be integrate out. In this situation, the Ricci scalar is found out to be
\begin{equation}
\mathcal{R} = \frac{6 \dot{t}}{a^{2} N^{4}}(a \ddot{a}\dot{t} - a\dot{a}\ddot{t} + N^{2}\dot{t}).
\end{equation}
  From (\ref{main_lag})  the Lagrangian takes the form in 4 dimensions as \cite{cordero1, BMP_RT}\footnote{Here $H^2=\frac{\Lambda}{3\beta}$ is a constant quantity.}
\begin{equation}
L(a, \dot{a}, \ddot{a}, \dot{t}, \ddot{t}) = \frac{a\dot{t}}{N^{3}} \left({a \ddot{a} \dot{t}-a \dot{a} \ddot{t} + N^{2}\dot{t}} \right) - N a^{3} H^{2}.
\label{hdlag}
\end{equation}
Note that the Lagrangian (\ref{hdlag}) contains higher derivative terms of the field $a$. However we can write it as \cite{cordero1}
\begin{equation}
L= -\frac{a{\dot {a}}^2}{N} + aN\left(1 - a^2H^2\right) + \frac{d}{d\tau}\left(\frac{a^2{\dot{a}}}{N}\right).
\label{orglag}
\end{equation}

It is worthy to note that the boundary term in this case plays no role in the equations of motion. However being a gravitational system, it is better to keep it at the first place as it carries important information about entropy. Also this will enable us to understand a HD gravity theory. For the Hamiltonian analysis of the system we will consider the HD version i.e. (\ref{hdlag}).  
 
\section{Hamiltonian analysis } tmp  

In this section we will perform the Hamiltonian formulation to extract the inherent symmetries of the higher derivative system (\ref{hdlag}). Particularly, we will follow the method of first order formalism developed in \cite{BMP}. This  is a different approach than the well known Ostrogradski method. We give   results rather shortly as details can be gathered from \cite{BMP_RT}.   

To begin with, we convert the HD Lagrangian (\ref{hdlag}) into a first order Lagrangian by defining  the time derivative of the fields as new fields. Thus in the equivalent first order formalism, we define the new  fields as,  
\begin{eqnarray}
\nonumber
\dot{a} &=& A,\\
\dot{t} &=& T.
\end{eqnarray}
These definitions  introduce new constraints in the system given by 
 \begin{eqnarray}
\nonumber
A -\dot{a} \approx 0, \\
 T - \dot{t}  \approx 0.
\label{hdconst}
\end{eqnarray}
Enforcing the constraints (\ref{hdconst})  through the Lagrange multipliers $\lambda_{a}$, and $\lambda_{t}$ as 
\begin{equation}
L^{\prime}=\frac{aT}{\left(T^{2}-A^{2}\right)^{\frac{3}{2}}}\left( {aT\dot{A}-aA\dot{T}+\left(T^{2}-A^{2}\right)T}\right) - \left( T^{2}-A^{2}\right) ^{\frac{1}{2}}a^{3}H^{2} + \lambda_{a}\left(A-\dot{a}\right) + \lambda_{t}\left(T-\dot{t}\right),
\label{auxlag} 
\end{equation}
we get the first order Lagrangian, known as the auxiliary Lagrangian. 
The Euler Lagrange equation of motion, obtained    from the first order Lagrangian (\ref{auxlag}),  by varying w.r.t. a, A, t, T, $\lambda_{a}$ and $\lambda_{t}$, are respectively given  by
\begin{eqnarray}
\frac{2a(\dot{A}T^{2} - AT\dot{T})}{(T^{2}-A^{2})^{\frac{3}{2}}} + \frac{T^{2}}{(T^{2}-A^{2})^{\frac{1}{2}}} -3 a^{2}H^{2}(T^{2}-A^{2})^{\frac{1}{2}} + \dot{\lambda}_{a} &=& 0,
  \label{lag_eom_a} \\
%\end{eqnarray}
%\begin{eqnarray}
 \frac{3a^{2}A(\dot{A}T^{2} - AT\dot{T})}{(T^{2}-A^{2})^{\frac{5}{2}}} - \frac{d}{d \tau}    \left( \frac{a^{2}T^{2}}{(T^{2}-A^{2})^{\frac{3}{2}}}  \right) - \frac{a^{2}T \dot{T}}{(T^{2}-A^{2})^{\frac{3}{2}}} + \frac{aAT^{2}}{(T^{2}-A^{2})^{\frac{3}{2}}} + \nonumber\\  \frac{a^{3}AH^{2}}{(T^{2}-A^{2})^{\frac{1}{2}}} + \lambda_{a} &=& 0, \label{lag_eom_A}\\ 
 %\end{eqnarray}
 %\begin{eqnarray}
\dot{\lambda}_{t} &=& 0, \label{lag_eom_t} 
\nonumber\\  
\frac{3a^{2}T(\dot{A}T^{2} - AT\dot{T})}{(T^{2}-A^{2})^{\frac{5}{2}}} + \frac{2a^{2} \dot{A}T}{(T^{2}-A^{2})^{\frac{3}{2}}} - \frac{d}{d \tau} \left( \frac{a^{2} A T}{(T^{2}-A^{2})^{\frac{3}{2}}}\right) -\frac{a^{2} A \dot{T}}{(T^{2}-A^{2})^{\frac{3}{2}}} + \frac{2aT}{(T^{2}-A^{2})^{\frac{1}{2}}} \nonumber\\ - \frac{aT^{3}}{(T^{2}-A^{2})^{\frac{1}{2}}} + \lambda_{t} &=& 0, \label{lag_eom_T}\\
A-\dot{a}&=&0, \label{lag_eom_lambda_a}\\
T-\dot{t}&=&0. \label{lag_eom_lambda_t}
\end{eqnarray} 
 To perform the Hamiltonian formulation we notice that the phase space is spanned by $q_{\mu}=a, t, 
 A, T, $ $\lambda_{a}, \lambda_{t}$ and their corresponding momenta  as $\Pi_{q_\mu}=\Pi_{a}, \Pi_{t}, 
 \Pi_{A}, \Pi_{T}, \Pi_{\lambda_{a}}, \Pi_{\lambda_{t}}$  with $\mu=0, 1, 2, 3, 4, 5$. Here the 
 momenta is defined by
\begin{equation}
\Pi_{q_{\mu}}  = \frac{\partial{L}^{\prime}}{\partial{\dot{q}_{\mu}}}.
\end{equation} 
 This is the point of departure of our Hamiltonian formulation from the Ostrogradsky convention of \cite{cordero1}.
 
Using the definition of momenta we find out the following set of primary constraints,
\begin{eqnarray}
\nonumber\\
\Phi_{1}&=& \Pi_{t} + \lambda_{t} \approx 0,
\nonumber\\
\Phi_{2}&=& \Pi_{a} + \lambda_{a} \approx 0,
\nonumber\\
\Phi_{3}&=&\Pi_{T} + \frac{a^{2}TA}{\left(T^{2}- A^{2}\right)^{\frac{3}{2}}} \approx 0,
\nonumber\\
\Phi_{4}&=& \Pi_{A} - \frac{a^{2}T^{2}}{\left(T^{2}-A^{2}\right)^{\frac{3}{2}}} \approx 0,
\nonumber\\
\Phi_{5}&=& \Pi_{\lambda_{t}}\approx 0,
\nonumber\\
\Phi_{6}&=& \Pi_{\lambda_{a}} \approx 0.
\end{eqnarray}
To obtain the full set of primary first class constraints we consider the constraint combination  
\begin{equation}
\Phi_{3}^{\prime} = T\Phi_{3} + A\Phi_{4} \approx 0.
\end{equation}
 Computing the Poisson brackets we find that only $\Phi_{3}^{\prime}$ gives the zero Poisson brackets with  all the constraints. The nonzero Poisson brackets between the newly defined primary set of constraints $\Phi_{1},\Phi_{2},\Phi_{3}^{\prime},\Phi_{4},\Phi_{5},\Phi_{6},$ become
 \begin{eqnarray}
\nonumber\\
\left\lbrace {\Phi_{1}, \Phi_{5}}\right\rbrace &=& 1,
\nonumber\\
\left\lbrace {\Phi_{2}, \Phi_{4}}\right\rbrace &=& \frac{2aT^{2}}{\left(T^{2}-A^{2}\right)^{\frac{3}{2}}},
\nonumber\\
\left\lbrace {\Phi_{2}, \Phi_{6}}\right\rbrace &=& 1.
\end{eqnarray}
Now, expression for the  canonical Hamiltonian is given by  
\begin{eqnarray}
\nonumber\\
H_{can} &=& -\frac{aT^{2}}{\left(T^{2}-A^{2}\right)^{\frac{1}{2}}} + \left(T^{2}-A^{2}\right)^{\frac{1}{2}} a^{3}H^{2} -\lambda_{a}A - \lambda_{t}T,
\end{eqnarray}
while the  total Hamiltonian becomes
 \begin{equation}
 H_{T} = H_{can} + \Lambda_{1}\Phi_{1} + \Lambda_{2}\Phi_{2} +  \Lambda_{3}\Phi_{3}^{\prime} + \Lambda_{4}\Phi_{4}+ \Lambda_{5}\Phi_{5} + \Lambda_{6}\Phi_{6}.
 \label{H_T} 
 \end{equation}
 Here $\Lambda_{1},\Lambda_{2}, \Lambda_{3}, \Lambda_{4}, \Lambda_{5}, \Lambda_{6}$ are  undetermined Lagrange multipliers.
  Demanding the time evolution of the constraints $\Phi_{1}$, $\Phi_{5}$, $\Phi_{6}$ to be zero  ($ \{\Phi_{i}, H_{T}\} \approx 0$) the following Lagrange multipliers get fixed
 \begin{eqnarray}
 \nonumber\\
 \Lambda_{5}&=& 0,
 \nonumber\\
\Lambda_{1} &=& T,
 \nonumber\\
\Lambda_{2} &=& A,
\end{eqnarray}
whereas, conservation of $\Phi_{2}$ gives the following condition between $\Lambda_{4}$ and $\Lambda_{6}$
\begin{equation}
\frac{T^{2}}{\left(T^{2}-A^{2}\right)^{\frac{1}{2}}} - 3a^{2}H^{2}\left(T^{2} - A^{2}\right)^{\frac{1}{2}} + \Lambda_{6} + \Lambda_{4} \frac{2aT^{2}}{\left(T^{2}-A^{2}\right)^{\frac{3}{2}}}=0.
\label{32}
\end{equation}
Time preservation of the constraint  $\Phi_{3}^{\prime}$ gives rise to the following secondary constraint
 \begin{equation}
 \Psi_{1} = \frac{aT^{2}}{\left(T^{2}-A^{2}\right)^{\frac{1}{2}}} - a^{3}H^{2} \left(T^{2}-A^{2}\right)^{\frac{1}{2}} + \lambda_{t}T + \lambda_{a}A \approx 0.
\end{equation} 
Likewise, $\Phi_{4}$ yields the following secondary constraint 
\begin{equation}
\Psi_{2}=\frac{aAT^{2}}{\left(T^{2}-A^{2}\right)^{\frac{3}{2}}} - \frac{a^{3}H^{2}A}{\left(T^{2}-A^{2}\right)^{\frac{1}{2}}} - \lambda_{a} \approx 0.
\end{equation}
Time preservation of  $\Psi_{1}$ trivially gives 0 = 0. A similar analysis involving $\Psi_{2}$ yields, on exploiting (\ref{32}),  

\begin{eqnarray}
\nonumber\\
\Lambda_{4} &=& - \frac{\left(T^{2} - 3a^{2}H^{2}\left(T^{2}-A^{2}\right)\right) \left(T^{2}-A^{2}\right)}{a\left(3T^{2} - a^{2}H^{2}\left(T^{2}-A^{2}\right)\right)},
\nonumber\\
\Lambda_{6} &=& - \frac{\left({ T^{2} - 3a^{2}H^{2}\left(T^{2}-A^{2}\right) }\right)  ( T^{2}-a^{2}H^{2}\left(T^{2}-A^{2}\right)^{\frac{1}{2}} )}{\left({T^{2}-A^{2}}\right)^{\frac{1}{2}} \left({3T^{2} -a^{2}H^{2} \left({T^{2}-A^{2}}\right)}\right)}.
\end{eqnarray}
The iterative procedure is thus closed and no more secondary constraints or other relations are generated.

Finally, after manipulation of the constraints we arrive to the following set of constraints
\begin{eqnarray}
\nonumber\\
F_{1}&=&\Phi_{3}^{\prime}  =T\Phi_{3} + A \Phi_{4} \approx 0,
\nonumber\\
 F_{2}&=&\Psi_{1}^{\prime}= \Psi_{1} - \Lambda_{4}\Phi_{4} \approx 0,
\nonumber\\
S_{1} &=&  \Phi_{4} \approx 0,
\nonumber\\
S_{2}&=& \Psi_{2}= \frac{aAT^{2}}{\left({T^{2}-A^{2}}\right)^{\frac{3}{2}}} - \frac{a^{3}AH^{2}}{\left(T^{2}-A^{2}\right)^{\frac{1}{2}}} + \Pi_{a} \approx 0.
\label{secondclass}
\end{eqnarray}
Here $\{F_1, F_2\}$ are the pair of first class constraints and  \{$S_{1}$, $S_{2}$\} are the pair of send class constraints. $F_1$ is the primary first class constraint which is consistent with the fact that there is only one undetermined multiplier in the theory. For removal of the second class constraints the Poisson brackets will be replaced by the Dirac brackets which is defined by
\begin{equation}
\{f,g\}_{D} = \{f,g\} -\{f, S_{i}\}\Delta_{ij}^{-1}\{S_{j},g\},
\label{dbdef}
\end{equation}
where we have taken
\begin{equation}
\Delta_{ij}=\{S_{i}, S_{j}\}= -\frac {aT^{2} \left({3T^{2}-a^{2}H^{2} \left({T^{2} -A^{2}}\right)}\right)}{\left(T^{2}-A^{2}\right)^{\frac{5}{2}}}  \epsilon_{ij},
\end{equation}
with $\epsilon_{12} = 1$ and $ i, j = 1, 2$. 

We calculate the Dirac brackets between the basic fields which are given below (only the nonzero brackets are listed)
\begin{eqnarray}
\nonumber\\
\left\lbrace {a,A}\right\rbrace_{D}&=& -\frac {\left(T^{2}-A^{2}\right)^{\frac{5}{2}}}{aT^{2} \left({3T^{2}-a^{2}H^{2} \left({T^{2} -A^{2}}\right)}\right)},
\nonumber\\
\left\lbrace {a, \Pi_{a}}\right\rbrace_{D}&=& \frac{T^{2}+ 2 A^{2} - a^{2}H^{2} \left(T^{2}-A^{2}\right)}{\left(3T^{2} - a^{2}H^{2}\left(T^{2} - A^{2}\right) \right)},
 \nonumber\\
\left\lbrace {a, \Pi_{A}}\right\rbrace_{D}&=& -\frac{3aA}{3T^{2} - a^{2}H^{2} \left(T^{2}-A^{2}\right)},
 \nonumber\\
\left\lbrace {a, \Pi_{T}}\right\rbrace_{D}&=& \frac{a\left(T^{2}+ 2A^{2}\right)}{T\left(3T^{2}-a^{2}H^{2}(T^{2}-A^{2})\right)},
\nonumber\\ 
\left\lbrace {t,\Pi_{t}}\right\rbrace_{D}&=&1, \nonumber\\
\left\lbrace {A, \Pi_{a}}\right\rbrace_{D}&=& - \frac{A \left(T^{2}-A^{2}\right) \left(T^{2}-3a^{2}H^{2} \left(T^{2}-A^{2}\right)\right)}{aT^{2}\left(3T^{2} - a^{2}H^{2} \left(T^{2}-A^{2}\right)\right)} ,
\nonumber\\
\left\lbrace {A, \Pi_{A}}\right\rbrace_{D}&=& \frac{2 \left(T^{2}-A^{2}\right)}{3T^{2} - a^{2}H^{2} \left(T^{2}-A^{2}\right)},
 \nonumber\\
 \left\lbrace{A, \Pi_{T}}\right\rbrace_{D}&=& \frac{A \left(T^{2} + 2A^{2} - a^{2}H^{2} \left(T^{2}-A^{2}\right) \right)}{T\left(3T^{2} - a^{2}H^{2} \left(T^{2}-A^{2}\right)\right)},
 \nonumber\\
\left\lbrace {T, \Pi_{T}}\right\rbrace_{D}&=& 1, \nonumber\\
\left\lbrace {\Pi_{a}, \Pi_{A}}\right\rbrace_{D}&=& -\frac{a \left(2T^{4}+A^{2}T^{2} + a^{2}H^{2}(T^{2}-A^{2})(9A^{2}-2T^{2})\right)}{ \left(T^{2}-A^{2}\right)^{\frac{3}{2}}(3T^{2} - a^{2}H^{2} \left(T^{2}-A^{2}\right)) },
 \nonumber\\
\left\lbrace {\Pi_{a}, \Pi_{T}}\right\rbrace_{D}&=& \frac{aA \left(T^{4}+2T^{2}A^{2} + a^{2}H^{2}(T^{2}-A^{2})(T^{2}+6A^{2})\right)}{T\left(T^{2}-A^{2}\right)^{\frac{3}{2}}(3T^{2} - a^{2}H^{2} \left(T^{2}-A^{2}\right) )},
 \nonumber\\
\left\lbrace {\Pi_{A}, \Pi_{T}}\right\rbrace_{D}&=& -\frac{a^{2}T \left(T^{2} +2A^{2} - a^{2}H^{2} (T^{2}-A^{2})\right)}{\left(T^{2}-A^{2}\right)^{\frac{3}{2}}(3T^{2} - a^{2}H^{2} \left(T^{2}-A^{2}\right))}.
\label{intdb} 
\end{eqnarray}
Due to the Dirac brackets all the second class constraints become strongly zero. So in this theory there is only two first class constraints. Now we proceed to show the gauge symmetry of the system.

\subsection{Construction of the gauge generator}
To construct the BRST symmetries, it is important to find out the gauge symmetries of the theory. So, we construct the gauge generator which  is the linear combination of all the first class constraints given by, 
\begin{equation}
 G= \epsilon_{1} F_{1} + \epsilon_{2} F_{2}.
 \label{gaugegenerator11}
 \end{equation}
 Here $\{\Phi_a\}$ is the whole set of  constraints and $\epsilon_{a}, a=1, 2$ are the gauge parameters. However not all the gauge parameters $\epsilon_{a}$ are independent. The independent gauge symmetries can be found out by applying the algorithm developed in \cite{rothe}. In addition, one should careful about finding out the independent symmetries since it is a HD theory. For the HD
  theories the  commutativity of gauge variation and time translation i.e.
\begin{eqnarray}
\delta q_{n,\alpha} - \frac{d}{dt}\delta{q}_{n,\alpha -1} = 0, \left(\alpha > 1 \right),
\label{varsgauge}
\end{eqnarray} 
where $q_{n,\alpha}$ denotes the $\alpha$-th order time derivative of $q$, can play very important role in finding out the independent gauge parameters \cite{BMP}.  Now, to find out the independent gauge parameters we use \cite{rothe} which gives us 

 \begin{equation}
 \epsilon_{1}= - \Lambda_{3}\epsilon_{2} - \dot{\epsilon}_{2}. 
 \label{parameterrelation1}
 \end{equation}
 So here  $\epsilon_{2}$ may be chosen as the independent gauge parameter. 
This means  that there is only one independent gauge transformation which essentially is in conformation with the fact that there is only one independent primary first class constraints.

 The gauge transformations of  the fields  are given by 
 \begin{eqnarray}
 \delta{a} &=& \{a, G\}_D = -  \epsilon_{2} A, \label{gaugetrans_a}\\
 \delta{t} &=& -  \epsilon_{2} T, \label{gaugetrans_t}\\
 \delta{A} &=&   \epsilon_{1} A  - \epsilon_{2} \frac{\left(T^{2} - 3a^{2}H^{2} \left(T^{2}-A^{2}\right)\right)\left(T^{2}-A^{2}\right)}{a\left(3T^{2} - a^{2}H^{2} \left(T^{2}-A^{2}\right)\right)},
  \label{gaugetrans_A}\\
 \delta{T} &=&   \epsilon_{1} T .
 \label{gaugetrans_T}
 \end{eqnarray}
 This completes our analysis  for finding out the gauge symmetries of the system. We have seen that the RT model has one independent primary first class constraint which is consistent with the fact that there is only one independent gauge symmetry. 
  \section{BRST quantization} 
  As we have seen that the RT model possesses  a redundant degrees of freedom and has been identified as the diffeomorphism symmetry in \cite{BMP_RT}.
  The principle of gauge invariance is essential to constructing a workable RT model. But it is   not feasible to perform a perturbative calculation   without first  fixing the gauge i.e. adding terms to the Lagrangian density of the action principle which  break the gauge symmetry  to suppress these  unphysical  degrees of freedom. 
  Since there are two first class constraints, so  we take two well motivated gauge fixing term 
 \begin{eqnarray}
\varphi_{1} &=& \sqrt{T^2 - A^2} -1 \approx 0,
\label{gauge1} 
\end{eqnarray}
and 
\begin{eqnarray}
\varphi_{2} &=& T - \alpha a \approx 0.
\label{gauge2}
\end{eqnarray}
The first gauge is the very well known cosmic gauge and the second gauge was proposed in \cite{BMP_RT}. Here $\alpha$ is some number with the condition $\alpha \neq H$. With these choice of gauges the first class constraints become second class. 
It is also known that such gauge fixing condition will induce ghost terms for the model
which play s an important role in the proof of unitarity of the theory.
Now we perform the BRST symmetry analysis of this model. 
    \subsection{Gauge fixing and Faddeev-Popov action}
  For simplicity we rename the gauge parameter $\epsilon_2$ as $\epsilon$. From 
  (\ref{parameterrelation1}) we can see that the other gauge parameter becomes $- \Lambda_{3}\epsilon - \dot{\epsilon}$.
  The above gauge conditions can be incorporated at quantum level by adding the
  following term to the classical action:
  \begin{eqnarray}
 L_{gf}&=&  \lambda (\sqrt{T^2-A^2}-1 +T - \alpha a), 
  \end{eqnarray}
 here both the first class constraints are incorporated by a single Lagrange multiplier $\lambda$ as there is only one independent gauge symmetry. 
   To construct the ghost term in the effective Lagrangian we first compute the gauge variation of the gauges $\varphi_1$ and $\varphi_2$. They are given by 
  \begin{eqnarray}
 \delta{\varphi}_1 &=&  -\sqrt{T^2-A^2}\Lambda_3 \epsilon - \sqrt{T^2-A^2}\dot{\epsilon}      - \frac{A\left(T^{2} - 3a^{2}H^{2} \left(T^{2}-A^{2}\right)\right)\sqrt{T^2-A^2}}{a\left(3T^{2} - a^{2}H^{2} \left(T^{2}-A^{2}\right)\right)} \epsilon,
\\ 
 \delta \varphi_2   &=& T (-\Lambda_3 \epsilon - \dot{\epsilon}) + \alpha    \epsilon A.
  \end{eqnarray}
     So we can easily calculate the ghost term for the effective Lagrangian as 
  \begin{eqnarray}
 \nonumber
 {L}_{gh} &=& \bar{c} (\frac{\partial}{\partial \epsilon}\delta{\varphi}_1 + \frac{\partial}{\partial \epsilon}\delta \varphi_2 )c
\\ 
& =&    -\bar{c}\sqrt{T^2-A^2}\Lambda_3 c  - \bar{c}\sqrt{T^2-A^2}\dot{c}     - \bar{c}\frac{A\left(T^{2} - 3a^{2}H^{2} \left(T^{2}-A^{2}\right)\right)\sqrt{T^2-A^2}}{a\left(3T^{2} - a^{2}H^{2} \left(T^{2}-A^{2}\right)\right)}c 
\nl
\ \ \ \ \ \ \ - \bar{c} T\Lambda_3 c - \bar{c} T \dot{c}+ \bar{c} \alpha    A  c,
\end{eqnarray}
where $c$ is a ghost field and $\bar c$ is an anti-ghost field.
These ghost and anti-ghost fields have a geometrical interpretation Maurer-Cartan 
forms on diffeomorphism. 
By adding the gauge fixing and ghost terms to the classical Lagrangian, 
the effective Lagrangian is then given by
   \begin{eqnarray}
 L_{eff}&=&  L+\lambda (\sqrt{T^2-A^2}-1 +T - \alpha a)  -\bar{c}\sqrt{T^2-A^2}\Lambda_3 c  - \bar{c}\sqrt{T^2-A^2}\dot{c}   \nonumber\\
 &  -& \bar{c}\frac{A\left(T^{2} - 3a^{2}H^{2} \left(T^{2}-A^{2}\right)\right)\sqrt{T^2-A^2}}{a\left(3T^{2} - a^{2}H^{2} \left(T^{2}-A^{2}\right)\right)}c 
  - \bar{c} T\Lambda_3 c - \bar{c} T \dot{c}+ \bar{c} \alpha    A  c.
 \end{eqnarray}
Following the structure of the gauge transformations (\ref{gaugetrans_a} - \ref{gaugetrans_T})   of the fields,  we are able to construct the  BRST transformations as
  \begin{eqnarray}
 \delta_b{a} &=&  -  c A \eta \label{BRST_a}\\
 \delta_b{t} &=& -  c T \eta\label{BRST_t},\\
 \delta_b{A} &=&   (-\Lambda_3 c - \dot{c}) A  \eta - c \frac{\left(T^{2} - 3a^{2}H^{2} \left(T^{2}-A^{2}\right)\right)\left(T^{2}-A^{2}\right)}{a\left(3T^{2} - a^{2}H^{2} \left(T^{2}-A^{2}\right)\right)} \eta,
  \label{BRST_A}\\
 \delta_b{T} &=&  (-\Lambda_3 c - \dot{c}) T \eta,
 \label{BRST_T} \\
 \delta_b{c} &=& 0, \ \delta_b{\lambda}=0, \ \delta_b{\bar{c}}=\lambda \eta, 
 \end{eqnarray} 
 which leaves the above effective action invariant. Here $\eta$  is the Grassmann parameter of transformation.
It is easy to check that the above transformation is  nilpotent, i.e. $\delta^2_b =0$.
This transformations can be used to compute the Ward identities which will yield the
relation between different Green's function. Here we are also able to
define the anti-BRST transformation just by replacing the ghost 
by anti-ghost field and vice-versa.
  \subsection{BRST charge calculation}
  In this subsection we derive the total BRST charge corresponding to the above BRST
  symmetry. Utilizing the Noether's formula,
   we calculate the BRST charges for the different fields as 
\begin{eqnarray}
\nonumber  
  Q_a &=& \frac{\partial \mathcal{L}}{\partial \dot{a}} \delta_b a \frac{1}{\eta} =  - \lambda_a (-  c A \eta)\frac{1}{\eta} = \lambda_a c A,
\nonumber  \\
  Q_t &=&  - \lambda_t (-  c T \eta)\frac{1}{\eta} = \lambda_t c T,  
\nonumber  \\
  Q_A &=& \frac{a^2 T^2}{(T^2 - a^2)^{\frac{3}{2}}} \Big( (-\Lambda_3 c - \dot{c}) A   - c \frac{\left(T^{2} - 3a^{2}H^{2} \left(T^{2}-A^{2}\right)\right)\left(T^{2}-A^{2}\right)}{a\left(3T^{2} - a^{2}H^{2} \left(T^{2}-A^{2}\right)\right)}  \Big),
\nonumber  \\
  Q_T &=&  - \frac{a^2 T^2 A}{(T^2 - a^2)^{\frac{3}{2}}}  (-\Lambda_3 c - \dot{c}), 
\nonumber  \\
  Q_{ c} &=& Q_{ \bar{c}} = 0. 
  \end{eqnarray}
  Total BRST charge, which is a hermitian operator, is given by
  \begin{eqnarray}
  \nonumber
  Q &=&
   \lambda_a c A + \lambda_t c T + \frac{a^2 T^2}{(T^2 - a^2)^{\frac{1}{2}}}  c \frac{\left(T^{2} - 3a^{2}H^{2} \left(T^{2}-A^{2}\right)\right)}{a\left(3T^{2} - a^{2}H^{2} \left(T^{2}-A^{2}\right)\right)}.  \label{BRST_charge}
  \end{eqnarray}
  This operator
  operator $Q^2$ which implements the BRST symmetry in quantum Hilbert space should be nilpotent which is evident from the above expression.
  In order to get probabilistic interpretation of the model we must project out
  all the physical states in the positive definite Hilbert space.
 Now, this charge ($Q$)  annihilates the physical states of the
total Hilbert space as follows: 
\begin{eqnarray}
 Q |\mbox{phys}\rangle =0,\label{kug}
\end{eqnarray}
which helps in defining the    physical states in total Hilbert space of the theory.

Similarly, it is also possible to compute the anti-BRST symmetry of this model
where the role of ghost fields will be changed by anti-ghost field.
The conserved charge for anti-BRST symmetry ($\bar Q$) must also satisfy the 
Kugo-Ojima condition:
\begin{eqnarray}
 \bar Q |\mbox{phys}\rangle =0.
\end{eqnarray}
The solution of this  equation will give us the quantum mechanical wave function of the universe. The BRST charge obtained in (\ref{BRST_charge})can be  mapped easily to the Wheeler DeWitt equation obtained in \cite{BMP_RT}. Here, in this paper, in the BRST approach we have not solved the first class constraints and hence the quantization is  actually done in the extended phase space. The corresponding  form of the WDW potential will also depend upon the variables of  extended phase space. Here we did  only the Lagrangian formulation to construct the BRST charges. Whereas. there exist another process namely the BFV \cite{BFV} formulation by which one can construct the Hamiltonian  of the model with  ghost fields in the extended phase space. In this approach the ghost field  and the lagrange multipliers are treated as dynamical variable and is a gauge independent way. The Hamiltonian thus obtained enables one to directly write down the Wheeler DeWitt equation and  the Wheeler DeWitt potential will be straight forward.  
  \section{Finite field-dependent BRST transformation}
  The purpose of this section is to study the extended BRST symmetry, known as FFBRST transformation, by
  making the transformation parameter finite and field dependent.
  Such transformations have been studied in various contexts with various important
  motivations. We try to build such formulation first time in RT based minisuperspace models.  We achieve this goal by  deriving first the methodology of FFBRST transformation, originally advocated in Ref. \cite{jm}, in a much elegant way.
Then, we discuss its illustration part.

We start with the consideration of the fields of RT model, written collectively as $\phi$, as a function of   parameter $\kappa: 0\leq \kappa\leq1$ in such a manner  that 
the original fields and finitely transformed fields are described by its extremum values.
Specifically,   $\phi(\tau, \kappa =0) =\phi (\tau)$ defines the original (non-transformed) fields, however,
   $\phi(\tau, \kappa =1) =\phi' (\tau)$ refers the finite field-dependent BRST transformed fields.
Now, fields transform under an
  infinitesimal field-dependent BRST transformation as   \cite{jm},
  \begin{eqnarray}
  \frac{d\phi(\tau, \kappa)}{d\kappa} =s_b \phi(\tau,\kappa) \Theta'[\phi (\tau,\kappa)].
  \end{eqnarray}
  Here $s_b \phi$ refers the Slavnov variation. To get the FFBRST transformation, we
  first integrate 
 the above equation w.r.t. $\kappa$  from $0$ to $\kappa$, leading to
  \begin{eqnarray}
  \phi (\tau,\kappa) =\phi (\tau,0) + s_b\phi (\tau, 0)\Theta [\phi(\tau, \kappa)],
  \end{eqnarray}
which at boundary ($\kappa =1$) yields the FFBRST transformation \cite{jm},
    \begin{eqnarray}
  \phi' (\tau) =\phi (\tau) + s_b\phi (\tau)\Theta [\phi(\tau)].
  \end{eqnarray}
  Here $\Theta [\phi]$ is a finite field-dependent parameter related to infinitesimal 
  version by $\Theta [\phi] =\int d\kappa \Theta' [\phi]$.
 The remarkable feature of  this FFBRST transformation is that this leaves the Faddeev-Popov action invariant, however, the functional measure is not, 
  leading to a non-trivial Jacobian.
We compute the Jacobian of path  integral measure under
 such FFBRST transformation by  following the similar procedure 
 (up to some good extent) as  discussed in \cite{jm}. The Jacobian of functional measure
 under an infinitesimal change is given by
 \begin{eqnarray}
 {\cal D}\phi   = J(\kappa) {\cal D}\phi (\kappa) = J(\kappa +d\kappa) {\cal D}\phi (\kappa +d\kappa),
 \end{eqnarray}
 which further reads
  \begin{eqnarray}
 \frac{J(\kappa)}{J(\kappa +d\kappa) }  = \sum_\phi\pm \frac{{\delta}\phi (\kappa +d\kappa)}{{\delta}\phi (\kappa)},\label{jaco}
 \end{eqnarray}
 where $\pm$ signs are considered according to
  the nature of the fields $\phi$. For instance, $+$ is used for bosonic
  fields and $-$ for Fermionic ones.
Upon Taylor expansion, the relation (\ref{jaco}) yields,
 \begin{eqnarray}
  -\frac{1}{J}\frac{dJ}{d\kappa} d\kappa =  d\kappa\int d\tau \sum_\phi\pm s_b\phi(\tau,\kappa) \frac{\delta\Theta'[\phi(\tau,\kappa)]}{\delta\phi(\tau,\kappa)}.\label{jac0}
 \end{eqnarray}
This further simplifies to
 \begin{eqnarray}
 \frac{d\ln J}{d\kappa} =-\int d\tau \sum_\phi\pm s_b\phi(\tau,\kappa) \frac{\delta\Theta'[\phi(\tau,\kappa)]}{\delta\phi(\tau,\kappa)}.
 \end{eqnarray}
Upon integration w.r.t $\kappa$ with its limiting values, this leads to following expression:
  \begin{eqnarray}
  \ln J   &=&-\int_0^1 d\kappa\int d\tau \sum_\phi\pm s_b\phi(\tau,\kappa) \frac{\delta\Theta'[\phi(\tau,\kappa)]}{\delta\phi(\tau,\kappa)},\nonumber\\
  &=&- \left(\int d^4x \sum_\phi\pm s_b\phi(\tau) \frac{\delta\Theta'[\phi(\tau)]}{\delta\phi(\tau)}\right).
 \end{eqnarray}
By exponentiating the above expression, we get the explicit form of the Jacobian of functional measure as follows:
   \begin{eqnarray}
  J   = {  \exp\left(-\int d\tau \sum_\phi\pm s_b\phi(\tau) \frac{\delta\Theta'[\phi(\tau)]}{\delta\phi(\tau)}\right)}.\label{jac}
 \end{eqnarray}
Due to this Jacobian  under FFBRST transformation,  an 
effective action $S[\phi]$ of the generating functional of the theory   undergoes to following 
change
 \begin{eqnarray}
 \int {\cal D}\phi'\ e^{iS[\phi']} =\int {\cal D}\phi \ e^{iS[\phi ]-\int d\tau \left(\sum_\phi\pm s_b\phi 
 \frac{\delta\Theta'}{\delta\phi }\right)},\label{gen}
 \end{eqnarray}
 where  $\phi'$ refers the transformed fields collectively.
Here we draw a conclusion that under the whole procedure, the effective action of the theory
 gets a precise modification in their original expression by an extra piece. 
We will notice that under  such an analysis the  theory does not change
 on physical ground. The resulting action depends on a arbitrary value of $\Theta'$.
In the next subsection, we illustrate this result by considering an specific value of 
$\Theta'$.
  \subsection*{Jacobian calculation}
For illustrating the above results discussed in the last section, we compute
the Jacobian of  functional measure explicitly.
  To compute the Jacobian  under a specific FFBRST transformation, 
  we first construct an infinitesimal field-dependent parameter with following form:
  \begin{eqnarray}
  \Theta'[\tau]=-i \bar{c}\left[\sqrt{T^2-A^2} -1 + T - \alpha a - F[\varphi]\right],\label{th}
  \end{eqnarray}
  where $F[\varphi]$ is an arbitrary gauge-fixing condition with $\varphi\equiv T,A,a,\alpha$.
 In construction of this parameter, we  take care of the ghost number of $\Theta'$, which must be $-1$.

  Now, the expression (\ref{jac}) together with (\ref{th}) yields
   \begin{eqnarray}
  J  & =&   \exp\left[i\left(-\lambda (\sqrt{T^2-A^2}-1 +T - \alpha a) +\bar{c}\sqrt{T^2-A^2}\Lambda_3 c  + \bar{c}\sqrt{T^2-A^2}\dot{c} \right. \right. \nonumber\\
 &  +&\left.\left. \bar{c}\frac{A\left(T^{2} - 3a^{2}H^{2} \left(T^{2}-A^{2}\right)\right)\sqrt{T^2-A^2}}{a\left(3T^{2} - a^{2}H^{2} \left(T^{2}-A^{2}\right)\right)}c 
  + \bar{c} T\Lambda_3 c + \bar{c} T \dot{c}-\bar{c} \alpha    A  c +\lambda F[\varphi] -\bar  
  c s_b \varphi \frac{\delta F}{\delta \varphi}\right)\right]. 
 \end{eqnarray}
 This corresponds to the Jacobian for
 path integral measure under FFBRST transformation with parameter in (\ref{th}).
 With this Jacobian, the functional integral (under FFBRST transformation) changes to
 \begin{eqnarray}
 \int {\cal D}\phi'\ e^{i\int d\tau L_{eff} [\phi']}& =&\int J[\phi]{\cal D}\phi \ e^{i\int d\tau L_{eff} [\phi]},\nonumber\\
 & =& \int {\cal D}\phi \ \exp\left[ i\int d\tau\left( L +\lambda F[\varphi]  -\bar  
  c s_b \varphi \frac{\delta F}{\delta \varphi} \right)\right],
 \end{eqnarray}
  which is nothing but the expression for generating functional of minisuperspace model in an
  arbitrary gauge. This assures that the FFBRST transformation
  maps the  minisuperspace model in one  specific gauge to any other arbitrary gauge.
  Since the gauge choice from one to another does not change the theory on physical ground
  as gauge-fixed action is BRST exact.
  Thus, we conclude that the FFBRST transformation  with any particular parameter does
  not change the theory numerically.

 \section{Conclusions}
 In this paper, we have analysed the BRST symmetries of the  Einstein-Hilbert(EH) action in the minisuperspace representation. Einsitein's theory includes first class constraints for which BRST symmetries play significant role during quantisation. We constrain  the generalised EH action, to 4+1 dimension. So the bulk we considered is 5 dimensional while the gravity is induced on the 4D hypersurface. The cosmological constant has been considered here for the reason that  five dimensional theory cannot have a vacuum solution\cite{kasner}. For the homogeneous and isotropic FRLW background, the Lagrangian which was obtained  appear to be as  pseudo higher derivative in nature. The Lagrangian we derived with respect to  the minisuperspace variables contains a total derivative term in addition to the  first order term. Without avoiding the surface term we considered the higher derivative version as existence of the surface term is directly linked to the thermodynamics of the system . Dirac's constraint analysis  showed that there are one primary first class constraint and one secondary first class constraint. We have seen that the  model is consist of only one gauge symmetry i.e. the reparametrisation invariance. The gauge symmetry is due to the existence of the primary first class constraints.   
   Looking at the gauge structure of the system, analogous BRST symmetries were constructed.  We considered the cosmic gauge and the gauge proposed in \cite{BMP_RT} to construct the effective Lagrangian. As there is only one  gauge symmetry these two constraints were added to the effective Lagrangian by a single Lagrange multiplier. The BRST transformations for all the fields in the extended phase space were calculated. Further we have calculated the BRST charges utilizing Noether theorem. The charge constructed so annihilate the physical states. The finite field-dependent BRST (FFBRST) transformations were also analyse for the theory.
   We have computed the Jacobian for the path integral  under FFBRST transformation with an arbitrary parameter which amounts a precise change in effective action.
   The results  were also illustrated with a particular construction of parameter
   which maps the action in a specific gauge to an arbitrary gauge.
  
     The aim to construct the BRST symmetries for a higher derivative system has been completed in the previous sections. Utility of the first order formalism to extract the gauge symmetries for the higher derivative system was invoked in the present paper. In fact, apart from just a mathematical tool the method described here can be helpful for  other higher dimensional models like the Randall-Sundurum background or more likely the Dvali-Gabadadze-Porrati extensions \cite{DGP}. As the Randall-Sundurum(RS)  model deals with the extra dimensions, one can intriguingly find  connection of the model discussed in this paper. The induced metric \ref{metric} which is  four dimensional here can give the results as obtained here only after consideration of the normal vector \ref{normal_vec}. So in the embedded space the choice of  normal vector is crucial. Consequently, a proper normal vector will  lead us to the metric discussed in \cite{DGP}.  One can further use these result to quantize the system and use the more tools like renormalization from quantum field theories and can see the outcomes. The analysis is left here with the scope for future consideration and as a step towards quantization in the path integral method for better understanding of the more complicated higher derivative  actions.

 %\section*{Acknowledgement}


\begin{thebibliography}{999}
\bibitem{stelle} K. S. Stelle, Phys. Rev. D \textbf{16} (1977) 953.
  
\bibitem{gibbon} G. W. Gibbons and S. W. Hawking, Phys. Rev. D 15(1977) 2752.
 \bibitem{regge} T. Regge and C. Teitelboim,  \textit{Proceedings of the Marcel Grossman Meeting}, Trieste, Italy, 1975, edited by R. Ruffini (North-Holland, Amsterdam, 1977), p. 77.

\bibitem{BMP_RT} R. Banerjee, P. Mukherjee and  B. Paul, Phys. Rev. D  \textbf{89} (2014) 043508.

  \bibitem{davidson1} A. Davidson, D. Karasik, and Y. Lederer, Class. Quant. Grav. \textbf{16}  (1999) 1349; Phys. Rev. D72, 064011 (2005).

  \bibitem{karasik} D. Karasik and A. Davidson, Phys. Re.v D \textbf{67 } (2003) 064012.
 
  
 
\bibitem{cordero1}  R. Cordero, A. Molgado and  E. Rojas, Phys. Rev. D \textbf{79} (2009) 024024.
\bibitem{cordero2}  R. Cordero, A. Molgado and  E. Rojas,  Gen. Relativ. Grav. \textbf{46} (2014) 1761 [arXiv:1309.3031].
\bibitem{BMP} R. Banerjee, P. Mukherjee and B. Paul, JHEP \textbf{08} (2011) 085.

\bibitem{BPU} R. Banerjee, B. Paul and  S. Upadhyay,  Phys. Rev. D \textbf{88}, 065019 (2013) [arXiv:1306.0744].
\bibitem{ht} M. Henneaux and C. Teitelboim,  \textit{Quantization of gauge
systems}  (Princeton, USA: Univ. Press, 1992).
\bibitem{wei} S. Weinberg,   \textit{The quantum theory of fields, Vol-II: Modern
applications} (Cambridge, UK Univ. Press, 1996).
\bibitem{hon} S.-T. Hong and Y.-J. Park, Phys. Rept. \textbf{358}  (2002) 143.
\bibitem{jm} S. D. Joglekar and B. P. Mandal,  Phys. Rev. D \textbf{ 51 } (1995)  1919.
\bibitem{ffb}   S. Upadhyay, Phys. Lett. B \textbf{740} (2015)  341;    Annls.  Phys. \textbf{340}  (2014) 110; Annls.  Phys.  \textbf{344} (2014)  290; EPL \textbf{105 } (2014) 21001;  EPL \textbf{ 104}   (2013) 61001; Phys. Lett. B \textbf{727 }  (2013) 293;  Annls.  Phys.  \textbf{356} (2015)   299; Mod. Phys. Lett. A \textbf{30} (2015) 1550072.

\bibitem{smm} S. Upadhyay, M. K. Dwivedi and B. P. Mandal,  Int. J. Mod. Phys. A 30 (2015) 1550178; Int. J. Mod. Phys. A 28 (2013) 1350033.
\bibitem{sudd} S. Upadhyay and D. Das, Phys. Lett. B \textbf{733 } (2014) 63.
\bibitem{rsu} R. Banerjee and  S. Upadhyay, Phys. Lett. B \textbf{734} (2014) 369.
 \bibitem{fsm}M. Faizal, S. Upadhyay and B. P. Mandal, Eur. Phys. J. C 76 (2016) 189; Phys. Lett. B \textbf{738} (2014) 201; Int. J. Mod. Phys. A \textbf{30 }  (2015) 1550032. 
\bibitem{sb} S. Upadhyay and B. P. Mandal,  Phys. Lett. B \textbf{744}  (2015) 231;
 	Eur. Phys. J. C \textbf{75} (2015) 327;  Int. J. Theor. Phys. \textbf{55} (2016)  1;   Prog. Theor. Exp. Phys.  \textbf{053B04} (2014) 1;  Eur. Phys. J.  {C \textbf{72}}   
(2012) 2065;  Annls.  Phys. \textbf{{327}} (2012) 2885; EPL  \textbf{{93}}  (2011)
{31001}; Mod. Phys. Lett.   {A  \textbf{25}}  (2010) {3347}.  

\bibitem{dirac} P. A. M. Dirac, Can. J. Math. {\bf 2} (1950) 129;  {\it Lectures on Quantum Mechanics}, Yeshiva University, 1964.

\bibitem{hanson} A.~Hanson, T.~Regge and C.~Tietelboim,  \textit {Constrained Hamiltonian System}, (Accademia Nazionale Dei Lincei, Roma, 1976).
 
 \bibitem{deser}  R. Arnowitt, S. Deser, C. W. Misner Gen. Relativ. Grav. \textbf{40} (2008) 1997.
 \bibitem{rothe}R.~Banerjee, H.~J.~Rothe and K.~D.~Rothe, Phys. Lett. B \textbf{463} (1999) 248 [hep-th/9906072]; Phys.~Lett.~B {\bf{ 479}} (2000) 429 [arXiv : hep-th/9907217]; R. ~Banerjee, H.~J. ~Rothe and  K.~D. ~Rothe,  J. Phys. A \textbf{33} (2000) 2059  [hep-th/9909039].
 \bibitem{kasner} Kasner, E., Am. J. Math. 43, 126 (1921).
 \bibitem{BFV} E. S. Fradkin and G. Vilkovisky, Phys. Lett. B \textbf{55} (1975) 224;
 I. A. Batalin and G. Vilkovisky, Phys. Lett.  B \textbf{69} (1977) 309;
 I. A. Batalin and E. S. fradkin, Phys. Lett.  B \textbf{122}  (1983) 157.
 \bibitem{DGP}  	L. Randall, R. Sundrum Phys.Rev.Lett. \textbf{83} (1999) 3370 [hep-ph/9905221], G. Dvali, G. Gabadadze, M. Porrati  	Phys.Lett.B \textbf{485} (2000) 208 [hep-th/0005016].
\end{thebibliography}
\end{document}